# DSNet: Dynamic skin deformation prediction by Recurrent Neural Network

1st Author
1st author's affiliation
1st author's E-mail address

2nd Author
2nd author's affiliation
2nd E-mail addresses

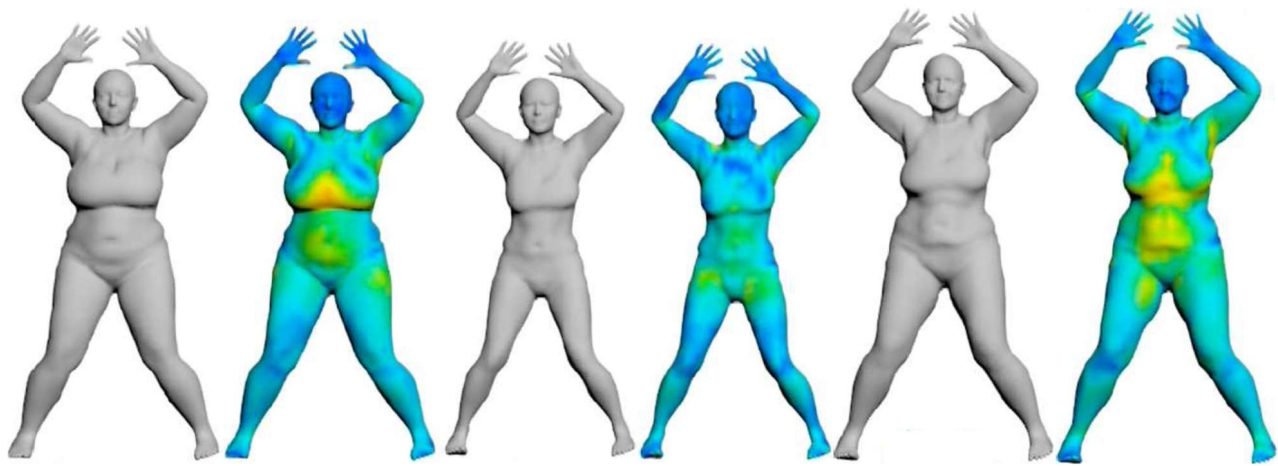

**Figure 1:** Our learning based dynamic skin modeler, DSNet takes a sequence of body meshes driven by skeleton motion as input and predicts dynamic skin deformation in a real-time course. The dynamic skin is learned as a function of pose, body identity shape, and the function values evaluated in previous frames. Such nonlinear skin dynamics contributes to the enhanced realism of human body.

## ABSTRACT

Skin dynamics contributes to the enriched realism of human body models in rendered scenes. Traditional methods rely on physics-based simulations to accurately reproduce the dynamic behavior of soft tissues. Due to the model complexity and thus the heavy computation, however, they do not directly offer practical solutions to domains where real-time performance is desirable. The quality shapes obtained by physics-based simulations are not fully exploited by example-based or more recent data-driven methods neither, with most of them having focused on the modeling of static skin shapes by leveraging quality data. To address these limitations, we present a learning-based method for dynamic skin deformation. At the core of our work is a recurrent neural network that learns to predict the nonlinear, dynamics-dependent shape change over time from pre-existing mesh deformation sequence data. Our network also learns to predict the variation of skin dynamics across different individuals with varying body shapes. After training the network delivers realistic, high-quality skin dynamics that is specific to a person in a real-time course. We obtain results that significantly saves the computational time, while maintaining comparable prediction quality compared to state-of-the-art results.

## 1. Introduction

Realistic skin or soft-tissue deformation for the human body model is a challenging task for many applications, which has a long tradition in computer graphics and computer animation. This is evidenced by the large amount of research done on geometric skin deformation, physics-based simulation, body shape capture and data-driven methods. The traditional skeleton driven skin deformation technique [34] and its variants are often widely practiced. Since these models mostly use linear transformations, it is difficult to model the nonlinear deformation such as bulging of muscles, not to mention the dynamic soft-tissue deformation on bodies under motion such as fat jiggling. These are crucial to capture and render the realism of a 3D body.

Physics-based models address the challenging problem of such nonlinear deformation and dynamics by adopting significantly more complex models [8, 40]. A common strategy is to model bones, muscles, and fats as volume elements (i.e. voxels), and to compute their time-varying deformation as the reaction to some external and internal forces by taking their physical properties into account. While these methods are well established in several domains like computer-aided medical intervention or computer aided design, and provide a possible solution to our problem, their model complexity and hence the heavy computation often make them impractical, especially for real-time applications.

Meanwhile, data driven methods have emerged as powerful models and seem to offer a good alternative. They seek for solutions to efficiently leverage realistic shapes coming either from interactive design by skilled designers, physically based simulation results, or captured shapes of real people. Earlier works rely on the sparse set of captured or manually elaborated data [37, 42], which have been gradually replaced by richer sets of captured shapes from real people, along with the advances in shape capture technology. Based on realistic shape datasets that are captured at high resolutions, mathematical models are learned to relate the model parameters to the generate detailed identity shapes and its deformation. Interestingly, data-driven models for the pose-dependent, nonlinear deformation [1, 2, 24, 27, 31, 39, 42] have been initially developed rather independently from the identity-dependent shape variation, whereas more recent models [3, 4, 17, 29, 36] incorporate both entities into a single, unifying framework. Many recent data-driven methods have shown to model static body shapes and pose-dependent shape deformations successfully. However, little attention has been paid to capturing and learning the dynamic skin- or soft-tissue deformations.

To address this problem, we pursue a new learning-based geometric model for the realistic dynamic skin deformation. The motivation behind our work is that the skin dynamics is highly nonlinear yet significantly limited by the kinematic constraints. Although the deformation of the muscle and fat driven by the bone motion is something complex to model physically, it can be approximated well by a simpler kinematic model as it boils down to some specific tasks in limited settings. Such an idea has been confirmed by the recent work Dyna [36], which has shown a successful application of data-driven approach to model such dynamics. There, the soft-tissue dynamics is learned as a function of the static body shape, a few inertia parameters (velocity and acceleration of joints), and the function values of previous frames. We share the idea of relying on the shape acquisition data to learn the parameters of kinematic approximator function on the dynamic shape deformation. Similarly to Dyna, our model bases its prediction on its previous predictions, and the velocities and accelerations of the body segments. We propose to deploy a deep neural network to learn the nonlinear approximator function for the dynamic skin behavior, instead of a linear function as in Dyna. A similar spirit has been shown in a recent work by Casas and Otaduy [10], who have trained neural network regressors to predict the nonlinear soft-tissue dynamics. However, our work distinguishes itself from theirs in that we model the dynamic skin deformation as a time-series entity, which we model by using a deep recurrent neural network. Moreover, our model can learn the subject-specificity of the skin dynamics, predicting a sequence of personalized skin deformation for each body. More recent work by Santesteban et al [38] has shown to approach the problem also by deploying a recurrent network. However, our model is simpler and lighter and thus more efficient to train and evaluate. Moreover, our model does not require any preprocessing of the input parameters, such as velocity or acceleration. Finally, the temporal scope of previous frames influencing the current frame is automatically learned as part of the network training, rather than predefined by using a constant.

At the core of our work is a deep neural network (DNN) to learn to approximate the detail and quality shapes of dynamic skin deformation from a dataset. An LSTM has been chosen due to its ability to learn over sequence data with its 'memory' on the history of its previous predictions. The shape encoder is based on a radial basis network to approximate the subject-specificity of the dynamic skin behavior from the dataset of only a few subjects. After the training, the network enables the real-time regression of nonlinear skin dynamics, which can be added to the linear blended skinning to obtain high quality skin deformation. We use captured body shapes from Dyna dataset, consisting of 10 subjects (5 female and 5 male), more than 120 motion clips of 3D body mesh sequence, each mesh containing more than 6000 vertices. To make the training feasible over such high-resolution mesh, we build a low-dimensional latent space to approximate the soft tissue deformation by using an autoencoder, over which the prediction network is trained.

Overall, our work makes the following contributions:
− DSNet that realistically predicts the nonlinear dynamics of skin deformation of a human body model under motion.
− A novel autoencoder that efficiently builds a low-dimensional latent space over the nonlinear skin offset of the body mesh, over which the predictor network is trained.
− A personalized skin dynamics simulation thanks to the learned function, producing distinct dynamic deformation according to the identity shape of the body.

## 2. Related work

Since the introduction of the Joint-dependent Local Deformation operators [34], the skeletal driven deformation or linear blend skinning (LBS) has become a well- established method for rigging a character. It involves first defining an underlying skeleton, associating each or a group of vertices on the skin mesh to their nearby bones, and deforming the skin mesh by applying to each skin vertex the weighted sum of the rotation transformations induced by its associated bones. Although it is easy to implement and efficient in computation, it has limited expressivity inherent to the linear model, with some noticeable artefacts such as volume loss, twisting ("candy-wrapper" effect), and self-intersection. Naturally, follow-up research has explored methods to solve these shortcomings.

**Geometric methods.** More advanced techniques propose using different representations of the rotation transformation. By replacing the transformation matrix with log-matrix blending [30], quaternion-based methods [16], dual quaternion [21], one can obtain improved skinning results efficiently while keeping the simplicity of linear combination.

Implicit skinning [46] and its extension [47] correct the geometric skinning results using implicit surfaces. They improve over the classical linear blending, allowing to circumvent the self-intersection artefacts and to model nonlinear deformation such as muscle bulging. However, such geometric models cannot model the soft-tissue dynamics such as jiggling of body parts, generated as a reaction to the rapid change of velocity and acceleration of a driving motion.

**Physics-based body models.** Physics-based methods can obtain highly realistic results by modeling the anatomical components of the body consisting of individual bones, muscles, and soft-tissues covered by a flexible skin. Since these are very complex structures that are not completely understood, various approximations have been proposed, such as early multi-layered geometric models [8], mass-spring models [35], or finite element models [44, 45]. Unfortunately, high-quality deformations come at the cost of model complexity and the additional computation of a hybrid or complete physical model, which is not quite suitable for real-time applications. Moreover, consisting of multiple entities for simulating the skeleton and internal organs, these models require accurate geometric models, and their creation is usually time-consuming. Performing such model creation independently for each subject makes such approaches impractical, hence many recent efforts have been devoted to the automatic generation of person-specific models [12]. Often, a template anatomical model containing the skeleton and internal organs is registered to a 3D surface scan by registration, through which the internal volume is adapted with constraints driven by the anatomical plausibility [9, 22].

**Example-based methods.** As an alternative to the LBS, example-based techniques have been developed, based on a sparse set of example skin shapes that were often generated by CG artists. Pose-space deformation methods [27, 42] formulate the skinning as a multi-way blending of several example skin shapes prepared at key skeletal poses (i.e., joint rotation angles). Given a new desired pose not seen, a plausible skin shape is approximated by interpolating example shapes with the similar poses.

Some other use auxiliary joints to best approximate example pose shapes with linear blend skinning. Mohr and Gleicher [31] removes LBS artefacts by adding an additional pseudo joint for each joint and computing vertex weights to best match the provided examples. This has been generalized by Kavan et al [20], allowing to add an arbitrary number of virtual bones instead of two, and whose optimal placement is automatically determined. Skinning decomposition models [11, 19, 26] extend this idea on animation sequences, trying to find the best approximating bone-vertex weights as well as bone transformations to the given example animation. While all these methods achieve realistic results with nonlinear muscle effects, they require considerable amount of tedious work from artists and do not model the skin dynamics.

**Data-driven body shape models.** With the rapid progression of computer and image sensing technologies, the analysis of deformable shapes captured by multi-view systems or 3D scanners has gained lots of interests. Not surprisingly, data-driven modeling of animated human bodies has been emerged as powerful methods in computer graphics since many years. They leverage the realistic shapes that are 3D scanned at high resolutions to parameterize the identity shape space of static body, and more recently, the space of both subject- and pose-dependent shapes. Early systems have focused on the generation of detailed identity shapes by learning mathematical models relating the model parameters to the shape linearly [2] or nonlinearly [39]. More recent models incorporate both subject- and pose-dependent shapes into a single, unifying framework. Based on hundreds of captured shapes of real people at dozens of key poses, SCAPE [4] or its more recent variant BlendSCAPE [17] allow the generation of new, posed body shapes at interactive time rate. They consider that an arbitrary body shape is represented in terms of triangle deformations applied to a template mesh, which are decomposed into identity-dependent shape and pose-dependent shape, which are modeled independently. SMPL (A Skinned Multi-Person Linear Model) [29] extends these earlier models by adopting a vertex-based skinning model and show that their model performs better than the deformation-based models. Although these models can achieve realistic skin deformation in interactive applications, little attention has been paid to capturing and learning the dynamic skin- or soft-tissue deformations.

More recently, with new multi-view systems enabling the acquisition of full 4D (3D+time) human shapes [7, 13, 23], data-driven methods that can capture, approximate, or model the skin dynamics have been proposed. MoSh [28] estimates the full body shape, pose, and soft-tissue deformations from the marker-based motion capture data, by fitting SCAPE parameters to the sparse marker set at each frame. While SCAPE does not explicitly model the soft-tissue deformation, soft tissue motions can still be captured by allowing the identity shape to vary across the sequence.

Closer to our work, Pons-Moll et al [36] have captured the dynamic human shape under movement using a high-resolution 4D capture system and learn a model of soft tissue deformations from those examples. They learn how soft tissue motion causes mesh triangles to deform relative to a base 3D body model using over 40,000 scans. Dyna model uses a low-dimensional linear subspace to approximate soft-tissue deformation and relates the subspace coefficients to the changing pose of the body. However, both their subspace and approximator function are linear, whereas the soft tissue motion is nonlinear by nature.

**Learning-based body models.** Bailey et al [6] use deep learning methods to model the nonlinear skin deformation function from an existing animation sequence. The mesh deformation is decomposed into a linear part that is computed directly from the transformations of the rig's underlying skeleton, and the remaining nonlinear part which is learned the network. The network after the training improves the speed while achieving a film quality deformation. However, the training is done per each subject and the nonlinear skin deformation is modeled as a function of static pose, omitting the dynamic aspect.

Our work is somewhat close to that of Casas and Otaduy [10], who proposed a neural network regressor that has been trained to predict the nonlinear soft-tissue dynamics. There are two main differences: First, we model the dynamic skin deformation as a temporal entity, i.e., temporal evolution is explicitly modeled, while in their work inter-frame dependency is only implicitly modeled through the input parameters. Second, our network learns the subject-specificity, producing personalized skin dynamics for each body. Compared to a more recent work [38] which also has deployed a recurrent network, our model is simpler and lighter and thus more efficient to train and evaluate. Moreover, both the inertia parameters (motion descriptor in [38]) and the temporal scope of previous frames influencing the current frame (2 frames in [38]) are automatically learned, as part of the network training.

**Recurrent neural network.** In this work we leverage the well-known capability of a recurrent network to capture temporal information, to model the temporal evolution of the nonlinear skin dynamics. In particular, LSTM [18] is a variant of RNN which include "memory" units that are specially designed to store information over long time periods, and thus can deal with long-term dependencies. It takes inputs, updates its internal state and its memory through recurrent connection that spans adjacent time steps, and generates outputs at every time-step iteratively. Therefore, the history of inputs affects the generation of outputs. LSTM has been successfully used in complex real-world sequence modeling tasks [14, 15].

## 3. Dynamic skin data and representation

We learn a dynamic skin deformation using a deep neural network. Such deep learning-based methods require large amount of dataset as training data. In our case, we train our network using the 4D dataset provided by Dyna [36]. AMASS (Archive of Motion Capture as Surface Shapes) [32] dataset, a large database of captured human shape under motion. It unifies different optical marker-based motion capture datasets by representing them within a common framework of SMPL model [29].

Our goal is to regress soft-tissue dynamics that, when added to existing blendshape models, will reproduce the nonlinear skin deformation effect. Thus, the shape model developed in SMPL is not sufficient and needs to be extended to incorporate an additional term to model the deformation caused by the skin dynamics. The use of such corrective blendshape for dynamics has been introduced in Dyna [36] model, which we use as base model in this work. We describe the representation model and the datasets below.

### 3.1 SMPL and DMPL models

SMPL models an arbitrary posed body shape as a linear blend skinning [34] of a subject-specific body surface that captures the characteristic of the body, given a set of subject-specific joint locations and the vertex-to-joint blending weights which are constant. The subject-specific body surface is obtained by applying per-vertex displacements to a predefined template model to capture the identity shape of the subject, and subsequent displacements to fix the artifacts of the linear blend skinning. The model is based on a linear subspace, which is found by the principle component analysis of thousands of 3D body scans coming from different subjects and poses. Formally, a surface body model $\mathbf{M}=M(\boldsymbol{\beta},\boldsymbol{\theta})$ is represented as

$$M(\boldsymbol{\beta},\boldsymbol{\theta}) = \mathcal{W}(\overline{M}(\boldsymbol{\beta},\boldsymbol{\theta}), J(\boldsymbol{\beta}), \boldsymbol{\theta}, \boldsymbol{W}) \quad (1)$$

$$\overline{M}(\boldsymbol{\beta},\boldsymbol{\theta}) = \overline{T} + M_S(\boldsymbol{\beta}) + M_P(\boldsymbol{\theta}), \quad (2)$$

where $\mathcal{W}(\boldsymbol{M}, \boldsymbol{J}, \boldsymbol{\theta}, \boldsymbol{W})$ is the linear blend skinning function that computes pose-dependent deformation of a body mesh $\boldsymbol{M}$ given the joint locations $\boldsymbol{J}$, the pose vector $\boldsymbol{\theta}$, and the vertex-to-joint blending weight matrix $\boldsymbol{W}$. $\boldsymbol{J}$ is a learned function to predict the joint locations from the subject-specific parameters $\boldsymbol{\beta}$, shape coefficient vector encoding the coordinates of an identity shape in the shape subspace learned from datasets.

The body surface in a rest pose $\overline{M}(\cdot)$ is obtained by adding vertex offsets computed by *shape blend shape* $M_S(\cdot)$, and subsequently adding those computed by *pose blend shape* $M_P(\cdot)$ to a learned rigged template mesh $\overline{T}$. This will capture the subject-specific shape, and correct the linear blend skinning artifacts, respectively. More specifically, $M_S(\cdot)$ is fully defined by the orthonormal principal components of shape displacements, $s_n \in \mathbb{R}^{3N}$, which is learned from registered training meshes. The body shape vector $\boldsymbol{\beta}$ encodes the coefficients to those principal components:

$$M_S(\boldsymbol{\beta}) = \boldsymbol{\mu}_S + \sum_{n=1}^{|\boldsymbol{\beta}|} \beta_n s_n.$$

Here $\boldsymbol{\mu}_S$ denotes the average shape displacement.

The pose blend shape function $M_P(\boldsymbol{\theta}): \mathbb{R}^{|\boldsymbol{\theta}|} \to \mathbb{R}^{3N}$ maps a pose vector $\boldsymbol{\theta}$ to a vector of vertex offsets by using vertex offset vectors $\boldsymbol{P}_n$, previously found from a dataset.

$$M_P(\boldsymbol{\theta}) = \sum_{n=1}^{9K}(R_n(\boldsymbol{\theta}) - (R_n(\boldsymbol{\theta}^*))\boldsymbol{P}_n,$$

where $K$ is the number of joints and $\boldsymbol{\theta}^*$ a rest pose. We refer [36] for details.

Dynamic SMPL, i.e., DMPL [36] extends SMPL formulation by adding *dynamic blend shape* $M_D(\cdot)$ to $\overline{M}(\cdot)$ in Eq. (2), another additive vertex offset accounting for soft-tissue dynamics:

$$M(\boldsymbol{\beta}, \boldsymbol{\theta}_t, \boldsymbol{\phi}_t) = \mathcal{W}(\overline{M}_t(\boldsymbol{\beta}, \boldsymbol{\theta}_t), J(\boldsymbol{\beta}), \boldsymbol{\theta}_t, \boldsymbol{W}) \quad (3)$$

$$\overline{M}_t(\boldsymbol{\beta}, \boldsymbol{\theta}_t, \boldsymbol{\phi}_t) = \overline{T} + M_S(\boldsymbol{\beta}) + M_P(\boldsymbol{\theta}_t) + M_D(\boldsymbol{\phi}_t, \boldsymbol{\beta}). \quad (4)$$

Here $\boldsymbol{\phi}_t = [\dot{\boldsymbol{\theta}}_t, \ddot{\boldsymbol{\theta}}_t, \boldsymbol{v}_t, \boldsymbol{a}_t, \boldsymbol{\delta}_{t-1}, \boldsymbol{\delta}_{t-2}]$ denotes the *dynamic control vector* at time *t*. It is composed of pose velocities and accelerations ($\dot{\boldsymbol{\theta}}_t$ and $\ddot{\boldsymbol{\theta}}_t$), root joint velocities and accelerations ($\boldsymbol{v}_t$ and $\boldsymbol{a}_t$), and the two *dynamic coefficient* vectors ($\boldsymbol{\delta}_{t-1}$ and $\boldsymbol{\delta}_{t-2}$) predicted in previous time steps. Note that the dynamic deformation $M_D(\cdot)$ is learned to predict the dynamic coefficient vector $\boldsymbol{\delta}_t$ at current time step, and is assumed to depend not only on the dynamic control vector but also on the shape characteristics inherent to a subject, such as fat distribution. Similarly to $M_S(\cdot)$, $M_D(\cdot)$ is represented by taking $|\boldsymbol{\delta}|$ principal components of soft-tissue displacements. Formally, it is written as:

$$M_D(\boldsymbol{\phi}_t, \boldsymbol{\beta}) = \boldsymbol{\mu}_D + \boldsymbol{D} \cdot f(\boldsymbol{\phi}_t, \boldsymbol{\beta}),$$

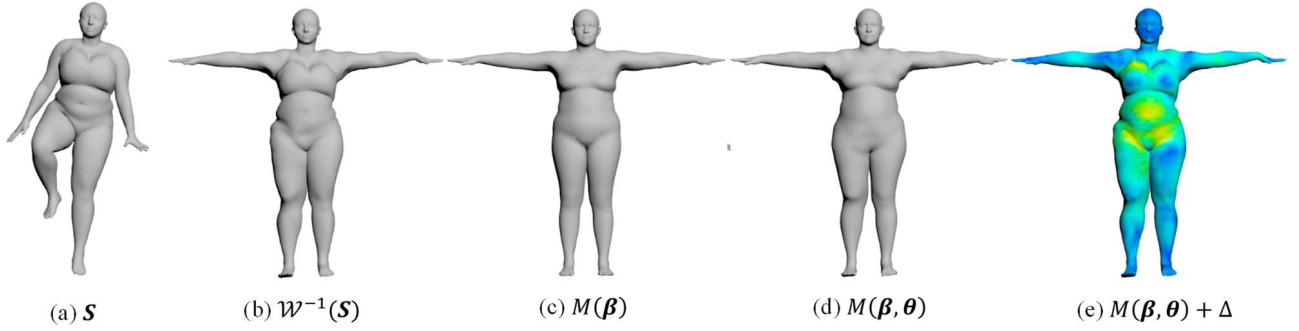

(a) $S$  (b) $\mathcal{W}^{-1}(S)$  (c) $M(\beta)$  (d) $M(\beta, \theta)$  (e) $M(\beta, \theta) + \Delta$

**Figure 2: The skin offset is found by computing the best (error-minimizing) alignment of the SMPL model $M(\beta, \theta)$ to the data mesh $S$. The residual deformation $\Delta = \mathcal{W}^{-1}(S) - M(\beta, \theta)$ as measured in the rest pose is considered as the contribution by the dynamic skin deformation.**

where $\mu_D$ denotes the average shape displacement, and $D = [d_n, ..., d_n] \in \mathbb{R}^{3N \times |\delta|}$ is the matrix of the principal components. $f(\cdot)$ is learned function that maps $\phi_t$ to $\delta_t$. Note that the soft-tissue displacements are the residual deformation computed in each frame by taking the difference between the SMPL model and the registered mesh, thus represent the dynamic deformations that depend on the body shape parameter $\beta$.

### 3.2 Datasets

To obtain a good level of accuracy and stable performance by the network, the training dataset needs to span a large range of motion and body shapes, which, unfortunately, is difficult to obtain. At the beginning we considered using a recent database of 4D captured shapes named AMASS (Archive of Mocap as Surface Shape) [32], a large database of human shape under motion unifying 15 existing motion capture datasets. Each frame in AMASS is represented as SMPL 3D shape identity parameters ($\in \mathbb{R}^{16}$), the DMPL soft tissue coefficients ($\in \mathbb{R}^8$), and the SMPL pose parameters ($\in \mathbb{R}^{159}$). However, not all motions in AMASS present dynamic skin deformation. In addition, the level of detail of the soft tissue deformation is rather limited, as only 8 parameters are made available. We therefore chose Dyna dataset [36] containing a variety of motions such as jumping and running in place, which allows to observe dynamic soft-tissue deformations.

As reviewed in Section 3.1, the body model from Dyna (Eq.(3), Eq.(4)) is based on a template triangular mesh $\bar{T}$ with $N=6890$ vertices that is rigged to a skeleton hierarchy with a total of $K_{Dyna}=22$ joints excluding hands. The template mesh is learned from a large scan dataset as a mean shape and is represented by a vector of $N$ concatenated vertex coordinates $\bar{T} \in R^{3N}$ in the rest (zero) pose $\theta^*$ and a set of blend weights, $W \in R^{N \times K}$. The skeletal structure is modeled by a kinematic chain connecting rigid bones by joints. Each joint has 3 rotational degrees of freedom parameterized with axis-angle representation, with an exception to the root joint that has 3 additional degrees of freedom of translation. Thus, the pose vector $\theta$ has 22×3+3=69 parameters. All subjects were all lightly clothed, while the motion duration varies across subjects and motions, from 2 to 15 seconds. Similarly to Dyna, we split male and female datasets to train networks separately. In this paper, we demonstrate the experimental results of our network trained with the female dataset.

To fully exploit Dyna dataset for the network training, and yet to be able to test the performance of the network on dynamics inducing motion, we used some of the skin-dynamics inducing motion sequences from Mosh [28] dataset, after some preprocessing to convert the frame rate (100 fps to 60 fps) and the orientation of the root joint (from *z*-up to *y*-up). Table 1 summarizes the number of subjects and of motions as well as their descriptions of two datasets used in this work.

**Table 1: Dyna and Mosh datasets we used in this work for training and validation, respectively. As part of AMASS [32] database, all triangular meshes are in a uniform representation of SMPL model.**

| dataset | subjects | motions | fps | No. sequences (men/women) |
|---|---|---|---|---|
| Dyna | 5 men, 5 women | 10~14 motions for each subject: one-leg jumping, light hoping, jumping jacks, shake hips, running in place, etc. | 60 | 66 / 67 |
| Mosh | Same subjects as above | Includes some skin-dynamics inducing motions (side-to-side hoping, basketball, kicking) that are not included Dyna. | 100 | 24 / 30 |

### 3.3 Generation of training data

Since we are interested in modeling the nonlinear *dynamic* skin deformation, we extract the vertex displacements contributed by that deformation ($M_D$ in Eq. (4)) from each mesh surface $S_t^j$ in the dataset, where $j$ is the index for the subject. (More accurately it is $S_t^{j,m}$ but we will omit the motion index $m$, for the sake of simplicity). We do so by finding the body shape vector $\beta^j$ that best matches the mesh at its first frame $S_t^j$, which is considered fixed throughout the rest of the motion sequence ($t=2,\ldots$). Formally, we compute $\boldsymbol{\beta}^j$ by solving:

$$\min_{\boldsymbol{\beta}^j, \boldsymbol{\theta_1}} \left\| \mathcal{W}\left(\overline{T} + M_S(\boldsymbol{\beta}^j) + M_P(\boldsymbol{\theta_1})\right) - S_1^j \right\|_2.$$

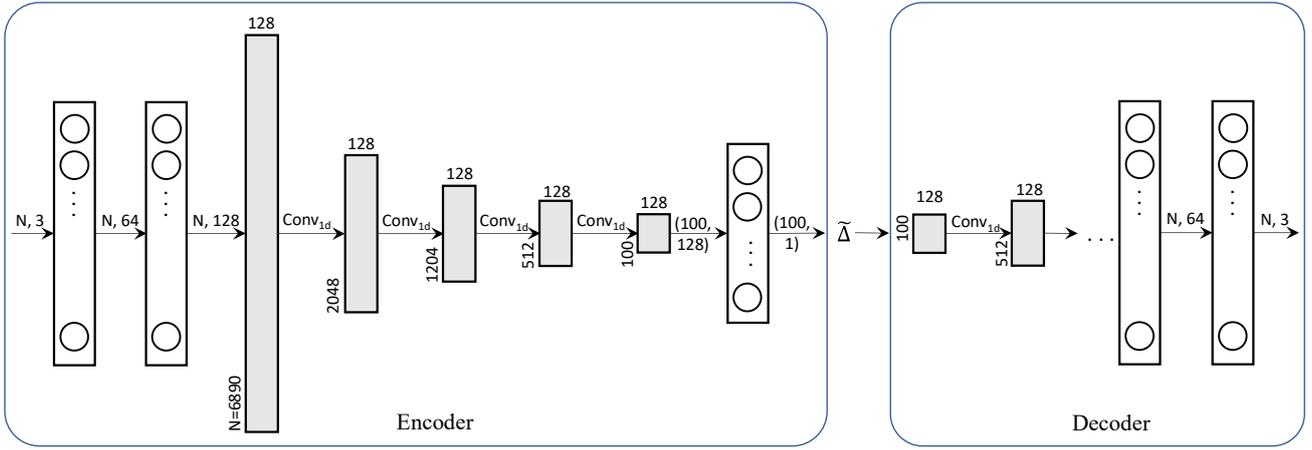

Figure 3: Our mesh autoencoder reduces the dimension of the original mesh 3N (3 × 6890 = 20,670) to a vector of 100 dimensions. The encoder is composed of two linear (fully connected) layers, followed by four 1-dimensional convolutional layers, and the last linear layer. The grey box represents the data, with its size denoted on top and bottom left. The decoder takes the symmetric structure of the encoder and reconstructs vertex coordinates of the mesh from the latent variable.

Next, for each subsequent frame, we compute the pose vector $\boldsymbol{\theta}_t$ that best aligns the model to the target mesh by solving:

$$\min_{\boldsymbol{\theta}_t} \left\| \mathcal{W}\left(\overline{T} + M_S(\boldsymbol{\beta}^j) + M_P(\boldsymbol{\theta}_t)\right) - S_t^j \right\|_2.$$

The dynamic skin offset $\Delta_t$ is the displacement vector originated from the dynamic skin deformation and is computed by:

$$\Delta_t = \mathcal{W}^{-1}(S_t) - \left(\overline{T} + M_S(\boldsymbol{\beta}^j) + M_P(\boldsymbol{\theta}_t^*)\right), \quad (5)$$

where $\mathcal{W}^{-1}$ denotes the *unposing* operation transforming a body mesh to its rest pose, and $\boldsymbol{\theta}_t^*$ denotes the best matching pose vector at frame $t$. Note that both the nonlinear *static* deformation $M_P$ and the dynamic deformation offset $\Delta$ is computed and added to body shape at its *rest pose*, similarly in [10]. We used stochastic gradient decent (SGD) for the parameter optimization with learning rate=0.0001 and momentum=0.9. Figure 2 shows some of the snapshots of mesh alignment results. We compute such dynamic skin offset $\Delta_t^m$ ($t=1,..T_m$) for all motions ($m=1,\ldots 67$) in the dataset. The training data is a set of input and output pairs ($\boldsymbol{\theta}_t^m, \Delta_t^m$), arranged in motion-major order. It is further processed: A dimension reduction of the dynamic skin offset vector by using an autoencoder (Section 4.1), and the uniformization of motion length (See Section 5) by tail-clipping or zero-padding.

## 4. DSNet

In this work we leverage the well-known capability of a recurrent network to capture temporal information, to model the temporal evolution of the nonlinear skin dynamics. At the core of our work is an LSTM network that learns to generate realistic dynamics-dependent deformations from observations. At runtime, our model takes as input a character undergoing a motion, whose skin mesh is deformed by skeletal driven deformation. As noted in Dyna, the dynamic deformations depend on body shape, the amount and the distribution of fat in the body, in particular. Thus, we train our DSNet such that it learns to generate dynamic shapes depending on the shape identity coefficients.

### 4.1 Dimension reduction by an autoencoder

An LSTM with hidden state of dimensionality $H$ takes feature vectors of dimensionality $d$ as input uses parameters in the order of $4(dH+d^2)$ [41]. The effectiveness of such models can be largely improved by reducing the dimensionality of the input data. Given the large number of data dimension (3×6890 = 20,670), it is desirable to reduce it into a compact, smaller sized latent representation. The majority of previous works use Principal Component Analysis, with an exception of [10] who use an autoencoder to make nonlinear dimension reduction. Here we also adopt an autoencoder, which compress the original data (tensor of size $N$ by 3 with vertex coordinates of a triangular mesh) to 100-dimensional latent space. The architecture of our autoencoder is shown in Figure 3. The first two fully connected layers gradually increases the feature description of each vertex. The subsequent CNN layers outputs lower resolution representations of the mesh, while keeping the feature dimension. The last linear layer learns to map the lowest resolution representation to a feature vector $\underline{\Delta}$ of 100 dimensions, which serves as the intrinsic representation of a body mesh. The decoder takes the symmetric structure of the encoder and reconstructs the original mesh data $\widetilde{\Delta}$ from the feature vector.

Some results of the qualitative evaluation of our autoencoder are shown in Figure 4. (We refer to the supplementary video for a complete visual validation.) The reconstruction error illustrated as the colormap is measured as the Euclidean distances between corresponding vertices. The trained network can faithfully reconstruct the detailed shape of the original mesh. When compared to the autoencoder network by Casas and

Otaduy, our network shows comparable performance, but it can be trained much more efficiently since it has much smaller number of parameters to be trained. The details on our autoencoder as well as the quantitative evaluation results can be found in Section 5 and Section 6.

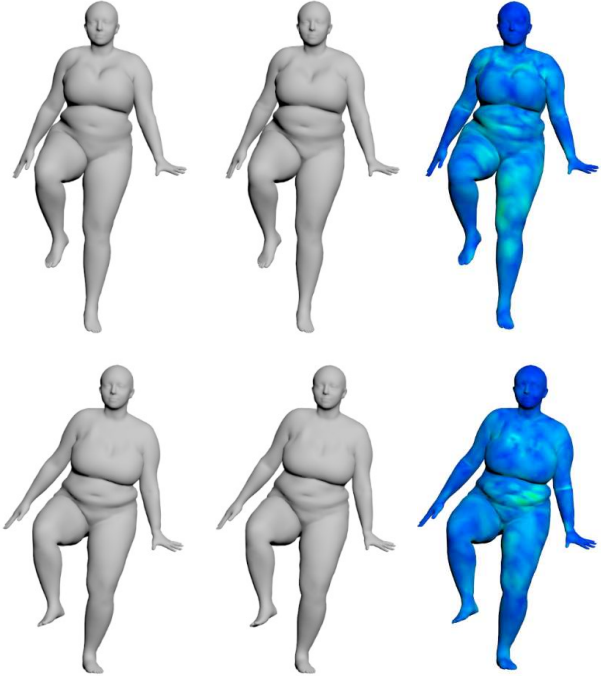

**Figure 4: The ground truth (the original mesh, left) and the network output (the reconstructed mesh, middle) from our autoencoder. The per-vertex error is illustrated as a color map (right).**

## 4.2 Learning dynamic skin deformation

Our goal is to learn a function that, given a sequence of kinematic pose vectors $\boldsymbol{\theta}_t$ ($t=1\ldots T$), predicts a sequence of dynamic skin offsets $\Delta_t$ ($t=1\ldots T$). Since what we want to predict is a time series, the results of each frame $t$ depend not only on the kinematic parameters encoded in the pose vector $\boldsymbol{\theta}_t$, but also on the results of previous frames $t$-1, $t$-2,… $t$-$l$. We should also consider the subject-specificity of the skin dynamics, which is related to the identity shape parameter $\boldsymbol{\beta}$. Formally,

$$\Delta_t = f(\boldsymbol{\theta}_t, f(\boldsymbol{\theta}_{t-1}); \boldsymbol{\beta}).$$

We model this with the long short term memory (LSTM) network, where the sequence of previous displacements is encoded as a fixed length hidden state memory. This memory is updated after seeing a new data (input and output pairs) by using non-linear functions. Note that this allows to learn an appropriate amount of history, i.e. how many previous frames to look back, to compute the skin offset in the current frame. This is contrary to the previous work where $l$ has set as a heuristically chosen constant (2 for example in [36]).

## 4.3 Dynamic skin deformation network

We employ the LSTM [18] as the basic architecture. LSTM is a variant of RNN which includes "memory" units that are specially designed to store information over long time periods. Since it preserves gradients well while backpropagating through time and layers, it can deal with long-term dependencies [43, 14, 15]. It takes inputs, updates its internal state and its memory through recurrent connection that spans adjacent time steps, and generates outputs at every time-step iteratively. Therefore, the history of inputs affects the generation of outputs.

As shown in Figure 5, our model is composed of a single layer of LSTM network, padded with two dense (fully connected) layers at the beginning and one dense layer at the end. Adding more LSTM layers did not improve the results, as it tends to widen its temporal scope and overfit easily with the increased number of layers. Initially we tried to build a radial basis function (RBF) network [5] based encoder to take the shape coefficient vector $\boldsymbol{\beta} \in \mathbf{R}^{10}$ as input and set the initial state of the LSTM network, considering that we have only 5 female subjects in Dyna dataset we use in this work, and that RBF network can learn a smooth approximation function from a sparse set of data samples. In this setting, the RBF based encoder will inform the decoder network about the subject-specificity of the target dynamic skin deformation in the form of an initial state vector for the hidden units. However, we obtained better results by informing the decoder network about the subject-specific shape at each time step, rather than once at the beginning. This is somewhat contrary to the work by Vinyals et al [48] on automatic image captioning, where the encoded image feature is fed into the decoder only once at the beginning.

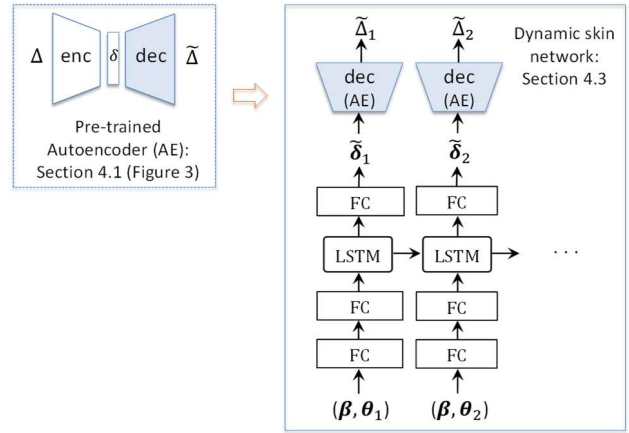

**Figure 5: The unrolled form of our Dyna-LSTM architecture.**

Our DS (Dynamic Skin) network takes the shape coefficient vector $\boldsymbol{\beta} \in \mathbf{R}^{10}$ concatenated with the kinematic pose vector $\boldsymbol{\theta}_t \in \mathbb{R}^{69}$ as input and learns to generate the kinematics-driven, subject-specific nonlinear dynamic skin deformation at each time step. The fully connected layers at the beginning learn to extract latent inertia features from $\boldsymbol{\theta}_t$, thus we do not need to compute the angular or linear velocity nor the acceleration unlike in Dyna [36]. The dense layer at the end maps the output from the LSTM network to an output vector $\widetilde{\boldsymbol{\delta}}_t \in \mathbf{R}^{100}$, a low dimensional feature vector in the latent space of dynamic skin deformations that has been pre-constructed by the autoencoder. We use the output vector $\widetilde{\boldsymbol{\delta}}_t$ to evaluate the autoencoder (decoder part) to reconstruct the

dynamic skin offset $\tilde{\Delta}_t$ to the body mesh in SMPL model. Formally it can be written as:

$$\tilde{\Delta}_t = AE_{dec}(\tilde{\boldsymbol{\delta}}_t), \quad t = 1,..T.$$

where

$$\tilde{\boldsymbol{\delta}}_t = FC_{enc\_3}(lstm_t),$$

$$lstm_t = LSTM\left(FC_{enc\_2}(FC_{enc\_1}(\boldsymbol{\theta}_t))\right).$$

**Loss function.** To find the optimal parameters of the network to predict the dynamic deformation coefficients, we define the loss as the sum of $L_2$-distances between the predicted dynamic coefficient vector $\tilde{\boldsymbol{\delta}}_t$ and the ground truth in the latent space $\boldsymbol{\delta}_t = AE_{enc}(\Delta_t)$ at each time step. The loss function L(·) is defined as:

$$L(\tilde{\boldsymbol{\delta}}_t, \boldsymbol{\delta}_t) = \sum_{t=1}^{T}\left(\|\tilde{\boldsymbol{\delta}}_t - \boldsymbol{\delta}_t\|_2\right).$$

The above loss is minimized with respect to all the parameters of LSTM- and the dense-layers through backpropagation through time [33]. The decoder LSTM layer shares the same parameters through time.

## 5. Implementation details

Our network models have been implemented using Tensorflow 2.0 (Dynamic skin deformation network) and Pytorch (mesh alignment, autoencoder) in a python environment. Experiments including the network training have been carried out on a regular desktop PC running on Ubuntu environment, with Nvidia GeForce 2080 Super graphics card.

**Data preprocessing.** Although spatially coherent (i.e. all meshes share a same topology), each motion sequence in Dyna database comes at different length thus different frame numbers. To feed in as input to the neural network, these sequence data need to be rearranged as a multidimensional tensor of a fixed shape, which means we need to uniformize the number of frames of all motion sequences. We heuristically chose a frame length, and zero-pad a sequence at the end if it is shorter or clip the tail out if it is longer. This frame length has been chosen based on the histogram (Figure 6) of motion sequences according to their original frame lengths. With most motion sequences having the frame lengths between 200 and 400, 300 has been chosen as the constant, as we want to avoid training the network with a dataset with too many 0's.

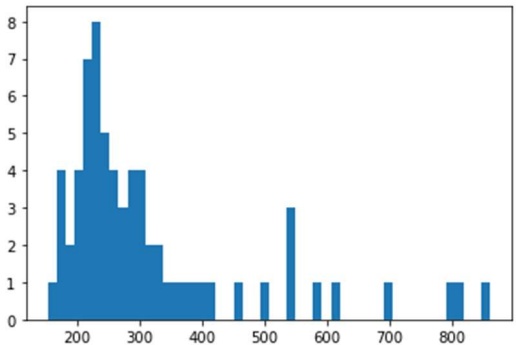

**Figure 6: Histogram of motion sequences in Dyna according to their original frame lengths.**

**Autoencoder.** All triangulated meshes available from Dyna dataset has been used. The training and validation sets consist of 15 051 (70%) and 6 451 (30%) meshes, respectively. The input data, i.e. the vertex coordinates have been normalized prior to input (so that their values vary within the range of [-1, 1]), and the network output has been denormalized. The network has been trained by using PyTorch implementation of the Adam optimizer [25], with a batch size of 64 and a learning rate of 0.0001. The total number of parameters are 33,626,606. Compared to the autoencoder from Casas and Otaduy [10], which has 3 fully connected layers of 6000, 3000, and 100 outputs each for the encoder, and the symmetric decoder, this amounts to only 11.8 % (=33,626,606 /284,678,770) of parameters to be trained, which has been confirmed by several hours of training time saved and much less memory requirements.

**Dynamic skin deformation network.** The first dense layer takes a 10-dimensional identity shape vector ($\boldsymbol{\beta} \in \mathbf{R}^{10}$) concatenated with a 69-dimensional pose vector ($\boldsymbol{\theta}_t \in \mathbb{R}^{69}$) as input and outputs a 64-dimensional vector, with a *linear* activation function. The second dense layer outputs a 128-dimensional vector with a *tanh* activation function, which is fed into the LSTM layer. The LSTM has 60 hidden units, and thus a 60-dimensional vector is produced as its output. This output vector goes through a batch normalization layer prior to the subsequent dense layer, which outputs a 100-dimensional vector $\tilde{\boldsymbol{\delta}}_t \in \mathbf{R}^{100}$.

**Training process.** The end-to-end training has been conducted using the motion sequences in the training dataset. The training time took approximately 0.05 seconds per epochs. The decoder network runs at high speed, since we use only 69 parameters for the pose parameter and 100-dimensional feature vector to encode the dynamic skin offset. Batch size was set to 16. Thus, data block of input and output training data is 16×300×69 (batch size, sequence length, pose parameter) and 16×300×100 (batch size, sequence length, feature vector), respectively.

We train the network via Tensorflow 2.0 implementation of Adam optimizer [25] with a learning rate of 0.0001. At each LSTM layer, a BatchNormalization layer has been appended so that their values vary within the range of [-1, 1]. We observed that it had a positive influence on the network performance in terms of error convergence.

## 6. Results

We have tested our DSNet to approximate the dynamic skin deformation on unseen subjects or/and motions. Whenever possible, the results are compared with the ground truth both quantitively and qualitatively. The dynamic skin offset (in centimeter unit) as measured by the per-vertex distances is illustrated in a color map throughout this section and in the accompanying video. Despite the rather limited size of training data, we have obtained very encouraging results. In particular, the network has learned to distinguish different body shapes, producing different dynamic skin deformations for each of them.

**Skin dynamics on validation data.** Figure 7 shows the results we obtained on a validation data, a motion sequence that has not been used for training. The chosen subject ('50004')'s other motions, as well as the semantically identical motions ('one leg jump') from other four subjects had been used for training the network. The predicted dynamic skin offset, when added to the SMPL model (Figure 7(d)), faithfully reproduces (Figure 7(b)) the original skin shape (Figure 7(a)). The temporal record of the error value (Figure 8) shows that error evenly distributed along the frames, with values ranging from 0.004 to 0.023.

**Skin dynamics prediction on unseen motions or unseen subjects.** Our DSNet can faithfully predict the skin dynamics for semantically new motions that have not been used for training. In Figure 9 and in the supplementary video, we visualize the dynamic skin deformation approximated for known and unknown subjects (framed in boxes) undergoing unobserved motions ('basketball dribble and shoot' in the middle rows, 'side to side hopping' in the bottom rows) by our DSNet. These results confirm that our DSNet has successfully learned to predict quality skin dynamics for unseen motions and/or unseen subjects.

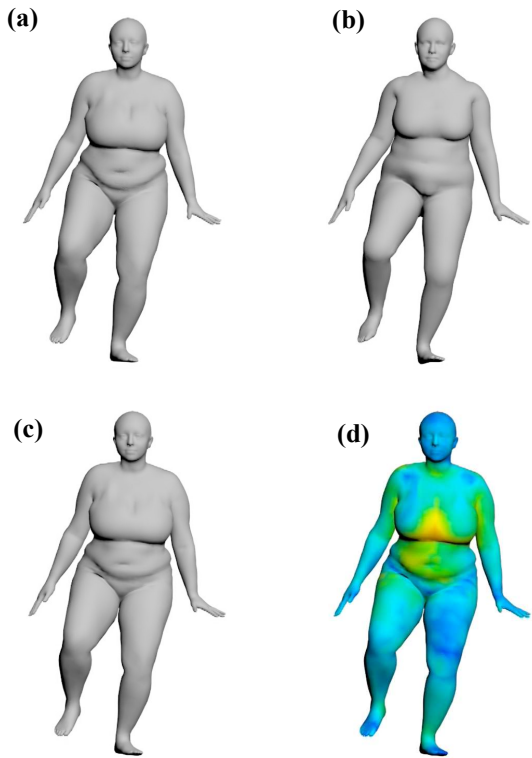

Figure 7: Dynamic skin deformation predicted for a validation data by DSNet. (a) Ground truth; (b) SMPL model reconstructed by using the input parameters (β, θ); (c) SMPL model appended with the predicted dynamic skin deformation; and (d) a color map indicating the amount of contribution by the DSNet to the final shape.

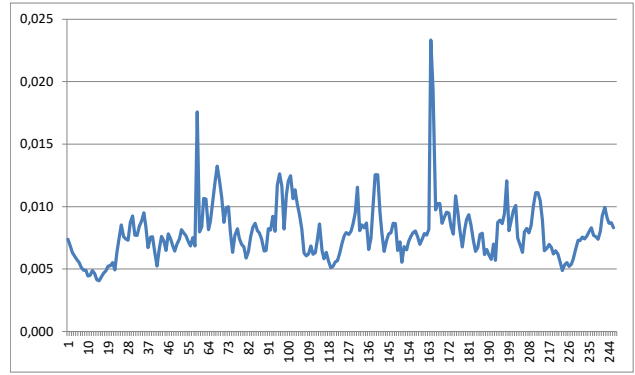

Figure 8: Temporal record of the error on 'one_leg jump' motion of subject '50004', which had been excluded from the training data.

Note that DSNet can also predict different dynamic skin deformation according to the varying shapes of different individuals. In Figure 9 and in the accompanying video, we can see that DSNet has produced distinct skin dynamics for each subject, even for an unknown one (framed in boxes).

**A note on the training data.** Ideally, the soft-tissue dynamics should be modeled solely by the dynamics-dependent shape ($M_D$ Eq.(4)), the training data for our neural network. However, we found that such dynamic shapes had been partly absorbed by the pose-dependent shape ($M_P$ Eq.(4)). As shown in Figure 10, the jiggling of breasts during 'jingle on toes' motion has been partly modeled by the oscillating rotation angle of 'spine2' joint (the spine joint near the breasts) during the per-frame based mesh aligning optimization. This is contrary to our visual inspection, which indicates the root and lower limbs are supposedly the only contributors to the jiggling deformation. This means that in our training data, the dynamics-dependent shapes separated by SMPL through optimization do not fully capture the observed dynamic shape. Although it can be compensated by *exaggerating* the predicted dynamic shape, it is something that can be improved.

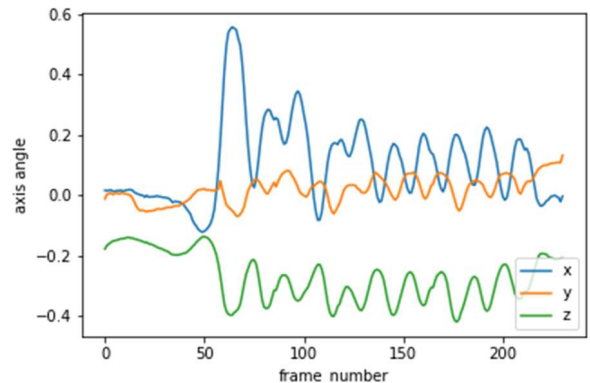

Figure 10. Rotation vector of 'spine2' joint in SMPL changes over time with 'jingle on toes' motion. (subject id: 50025) The dynamics of bouncing breast has been partly modeled the rotation of the spine joint.

# 7. CONCLUSION

We have presented a learning-based method to the quality estimation of dynamic skin deformation. The dynamic skin deformation has been modeled as a time series data, which is learned as a function of kinematic pose parameter, characteristics of body shape, and of the results of previous time steps. An LSTM-based network has been developed, which has been trained on sequences of high-quality triangular meshes captured from real people under diverse motions. Borrowing the SMPL human body shape representation, the dynamics-dependent skin deformation has been extracted from each frame mesh along with the pose parameters, which has been used as training data. Once trained, the network successfully predicts the nonlinear, dynamics-dependent shape changes over time, contributing to a high-quality skin dynamic of human body models under motion in a real-time course. Also developed has been an autoencoder, which builds a compact space for the intrinsic representation of dynamic skin offset and thereby allowing a very efficient operation of the dynamic skin deformation network. We have evaluated our model on various unseen motions and different individual shapes and shown that our model significantly saves the computational time and produces quality soft-tissue dynamics in real-time.

# ACKNOWLEDGMENTS

Omitted for the anonymous review.

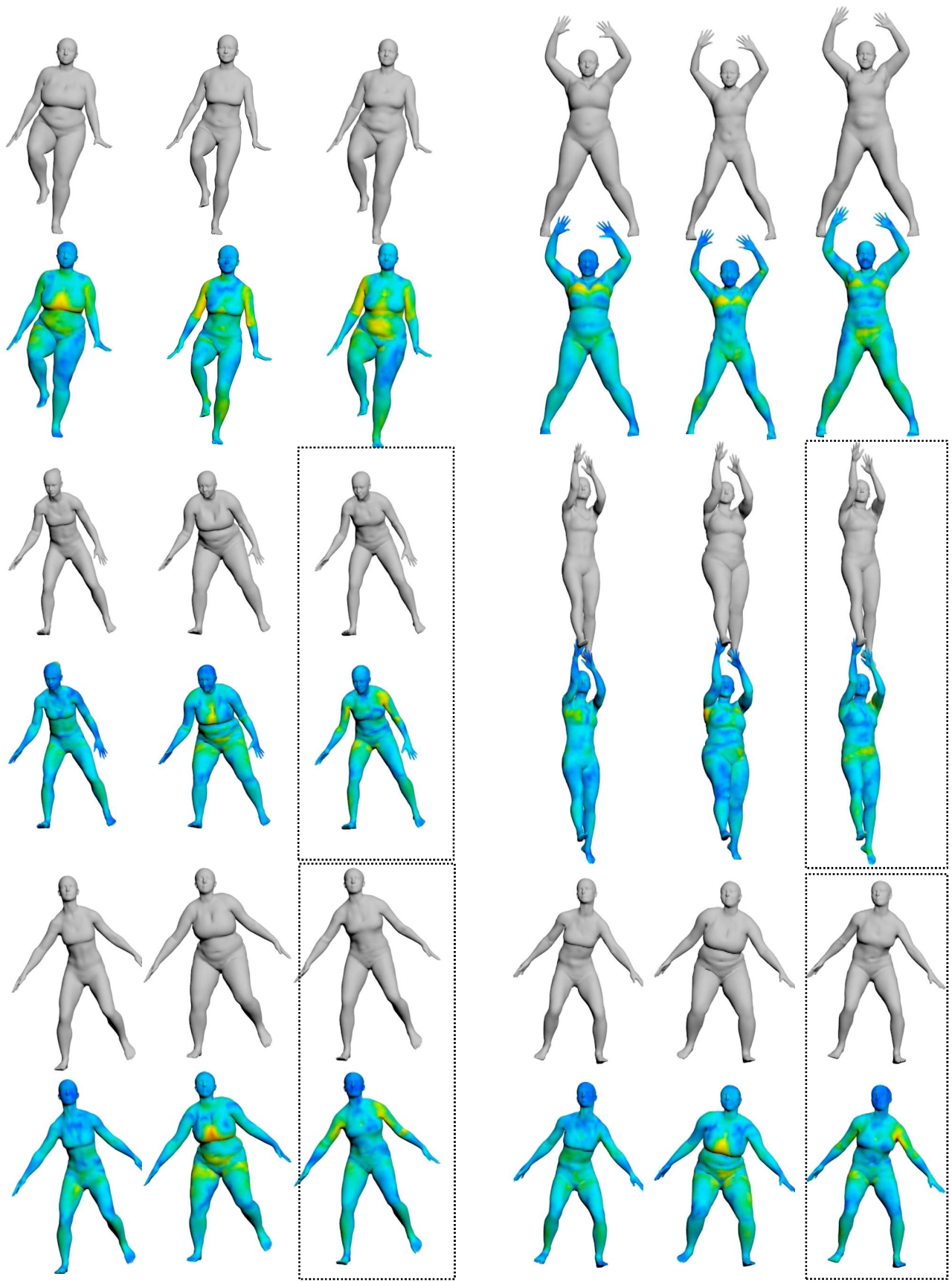

Figure 9: Dynamic skin deformation approximated by our DSNet for known and unknown (framed in boxes) subjects undergoing observed ('one leg jump':top left, 'jumping jaks':top right) and unobserved motions ('basketball dribble and shoot':middle rows, and 'side to side hopping': bottom rows).